\documentstyle[prd,eqsecnum,aps,epsf]{revtex}

\tightenlines
\begin{document}

%============ The new commands: =============================
    \newcommand{\DSC}{D\hspace{-0.25cm}\slash_{\bot}}
    \newcommand{\DSP}{D\hspace{-0.25cm}\slash_{\|}}
    \newcommand{\DS}{D\hspace{-0.25cm}\slash}
    \newcommand{\DC}{D_{\bot}}
    \newcommand{\DSCX}{D\hspace{-0.20cm}\slash_{\bot}}
    \newcommand{\DSPX}{D\hspace{-0.20cm}\slash_{\|}}
    \newcommand{\DP}{D_{\|}}
    \newcommand{\QV}{Q_v^{+}}
    \newcommand{\QVB}{\bar{Q}_v^{+}}
    \newcommand{\QVP}{Q^{\prime +}_{v^{\prime}} }
    \newcommand{\QVBP}{\bar{Q}^{\prime +}_{v^{\prime}} }
    \newcommand{\QVHZ}{\hat{Q}^{+}_v}
    \newcommand{\QVHZB}{\bar{\hat{Q}}_v{\vspace{-0.3cm}\hspace{-0.2cm}{^{+}} } }
    \newcommand{\QVPHZB}{\bar{\hat{Q}}_{v^{\prime}}{\vspace{-0.3cm}\hspace{-0.2cm}{^{\prime +}}} }

    \newcommand{\QVPHFB}{\bar{\hat{Q}}_{v^{\prime}}{\vspace{-0.3cm}\hspace{-0.2cm}{^{\prime -}} } }
    \newcommand{\QVPHB}{\bar{\hat{Q}}_{v^{\prime}}{\vspace{-0.3cm}\hspace{-0.2cm}{^{\prime}} }   }

    \newcommand{\QVHF}{\hat{Q}^{-}_v}
    \newcommand{\QVHFB}{\bar{\hat{Q}}_v{\vspace{-0.3cm}\hspace{-0.2cm}{^{-}} }}
    \newcommand{\QVH}{\hat{Q}_v}
    \newcommand{\QVHB}{\bar{\hat{Q}}_v}
    \newcommand{\VS}{v\hspace{-0.2cm}\slash}
    \newcommand{\MQ}{m_{Q}}
    \newcommand{\MQP}{m_{Q^{\prime}}}
    \newcommand{\QVHPMB}{\bar{\hat{Q}}_v{\vspace{-0.3cm}\hspace{-0.2cm}{^{\pm}} }}
    \newcommand{\QVHMPB}{\bar{\hat{Q}}_v{\vspace{-0.3cm}\hspace{-0.2cm}{^{\mp}} }  }
    \newcommand{\QVHPM}{\hat{Q}^{\pm}_v}
    \newcommand{\QVHMP}{\hat{Q}^{\mp}_v}

\newcommand{\PP}{{1 + v\hspace{-0.2cm}\slash \over 2}}
\newcommand{\PM}{{1 - v\hspace{-0.2cm}\slash \over 2}}

\newcommand{\CDPL}{\overleftarrow{\hat{\cal D}}\hspace{-0.29cm}\slash_v}
\newcommand{\CDPN}{{\cal D}\hspace{-0.26cm}\slash_v}
\newcommand{\CDPNL}{\overleftarrow{\cal D}\hspace{-0.29cm}\slash_v}
\newcommand{\DSCXL}{\overleftarrow{D}\hspace{-0.29cm}\slash_{\bot}}
\newcommand{\DSXL}{\overleftarrow{D}\hspace{-0.29cm}\slash}

\newcommand{\oDbot}{\overleftarrow{/\!\!\!\!D}\hspace{-5pt}_{\bot}}
\newcommand{\oDSP}{\not\!{v} v\cdot \overleftarrow{D}}
\newcommand{\odsp}{\overleftarrow{D}\hspace{-0.29cm}\slash_{\|}}

 %==============================================================

\draft
\title{$1/m_Q$ corrections to $B\to \rho l\nu$ decay and $|V_{ub}|$}
\author{W. Y. Wang\footnote{E-mail address: wangwenyu@tsinghua.org.cn}, Y. L. Wu, M. Zhong}
\address{Institute of Theoretical Physics, Academia Sinica,
 Beijing 100080, China }
\maketitle

\begin{abstract}
In the heavy quark effective field theory of QCD, we analyze the
order $1/m_Q$ contributions to heavy to light vector decays. Light
cone sum rule method is applied with including the effects of
$1/m_Q$ order corrections. We then extract $|V_{ub}|$ from $B\to
\rho l\nu$ decay up to order of $1/m_Q$ corrections.
\end{abstract}

\vspace{0.5cm}

\pacs{PACS: 11.55.Hx, 12.39.Hg, 13.20.Fc, 13.20.He
\\
 Keywords:
 $B \to \rho l\nu$, $1/m_Q$ correction,
 heavy quark effective field theory, light cone sum rule
}

\section{Introduction}\label{int}
Much effort has been devoted to discuss the heavy to light hadron
semileptonic decays. In particular, $B\to \pi (\rho) l\nu$ decays
attracted the most interest
\cite{var,ar,arsc,pvda,pvmisu,pvesrb,myc60,cs45,myc9594} because
they can be used to determine the quark mixing matrix element
$|V_{ub}|$, a parameter of significance in particle physics. The
heavy quark symmetry and relevant effective theory greatly
simplify the study of hadrons each of which containing a single
heavy quark and any number of light quarks, and provide relations
between different processes. This symmetry is applied to study $B
(D)_{(s)}\to \pi (\rho, K, K^*) l\nu$ decays in
Refs.\cite{bpi,brho,h2l}, where the finite heavy quark mass
($m_Q$) corrections are not considered. Ref.\cite{bpinlo} extends
the study on $B\to \pi l\nu$ decay up to the next to leading order
of the heavy quark expansion. For a more complete knowledge of the
magnitude of the finite mass corrections to heavy to light meson
decays, and to the determination of $|V_{ub}|$, one should also
study the $1/m_Q$ order corrections to semileptonic $B$ decays to
light vector mesons.

In this short letter we will apply the heavy quark effective field
theory (HQEFT) developed in Refs. \cite{ylw,wwy,yww,ww,excit} to
analyze the $1/m_Q$ corrections to the $B\to \rho l\nu$ decay. And
the light cone sum rule method will be adopted to numerically
estimate the nonperturbative functions, i.e., the heavy to light
vector form factors with including $1/m_Q$ order corrections. In
section \ref{formulation} the $1/m_Q$ order corrections are
formulated in HQEFT framework. Section \ref{sumrule} devotes to
evaluate wave functions using light cone sum rule method in HQEFT.
And section \ref{discussion} is the numerical results and
discussion.

\section{$B\to \rho l\nu$ decay in HQEFT}\label{formulation}

The transition matrix element responsible to the $B\to \rho l\nu$
decay is generally parameterized by form factors as
\begin{eqnarray}
\label{fdef} && <\rho(p,\epsilon^*)|\bar u\gamma^\mu (1-\gamma^5)
b|B(p_B)>=-i (m_B+m_\rho) A_1(q^2)
     \epsilon^{*\mu} +i \frac{A_2(q^2)}{m_B+m_\rho} (\epsilon^{*}\cdot (p+q) )
    (2p+q)^\mu \nonumber\\
&& \hspace{2cm}+i\frac{A_3(q^2)}{m_B+m_\rho} (\epsilon^* \cdot
(p+q)) q^\mu
   + \frac{2 V(q^2)}{m_B+m_\rho} \epsilon^{\mu \alpha \beta \gamma}
    \epsilon^*_\alpha (p+q)_\beta p_\gamma,
\end{eqnarray}
where $q=p_B-p$ is the momentum carried by the lepton pair.

In the framework of HQEFT \cite{ylw,wwy}, the QCD quantum field
$Q$ for heavy quark is decomposed into ¡°particle field¡± $Q^+$
and ¡°antiparticle field¡± $Q^-$, so that the quark and antiquark
fields are treated on the same footing in a symmetric way. The
effective quark and antiquark fields in HQEFT are defined as
\newcommand{\VSS}{v\hspace{-0.15cm}\slash}
\begin{eqnarray}
&&Q^{\pm}_v=e^{i\VSS m_Q v\cdot x} \hat{Q}^{\pm}_v =e^{i\VSS m_Q
v\cdot x} P_{\pm} Q^{\pm} \\
&&R^{\pm}_v=P_{\mp} Q^{\pm}
\end{eqnarray}
with $v$ being an arbitrary four-velocity satisfying $v^2=1$, and
$P_{\pm} \equiv (1 \pm \VS)/2$ being the projection operators.
$\hat{Q}^{\pm}_v$ defined above are actually the large components
of the heavy quark and antiquark fields respectively. $R^{\pm}_v$
are the small components of the heavy quark and antiquark fields
respectively. The quantum field in QCD Lagrangian can be written
as $Q\equiv Q^+ +Q^- \equiv
\hat{Q}^{+}_v+\hat{Q}^{-}_v+R^{+}_v+R^{-}_v$, which contains all
large and small components of particle and antiparticle. The
decomposition of $Q$ is presented in detail in
Refs.\cite{ylw,wwy}.

After the small components of particle and antiparticle fields
being integrated out, QCD Lagrangian turns into
\begin{eqnarray}
\label{HQEFTLagrangian} {\cal L}_{Q,v}={\cal L}^{(++)}_{Q,v}+{\cal
L}^{(--)}_{Q,v}+{\cal L}^{(+-)}_{Q,v}+{\cal L}^{(-+)}_{Q,v}
\end{eqnarray}
with
\begin{eqnarray}
    \label{lzz2}
{\cal L}^{(\pm \pm)}_{Q,v} &=&
        \bar{Q}^{\pm}_v i\CDPN Q^{\pm}_v , \\
 \label{lzf2}
{\cal L}^{(\pm \mp)}_{Q,v} & = &
              \frac{1}{2\MQ} \bar{Q}^{\pm}_v (i\CDPNL)  e^{2iv\hspace{-0.15cm}\slash m_Q v\cdot x}
        \Big(1-\frac{i \VS v\cdot D}{2\MQ}\Big)^{-1} (i\DSCX ) Q^{\mp}_v \nonumber \\
        & = & \frac{1}{2\MQ} \bar{Q}^{\pm}_v (-i\DSCXL) \Big(1-\frac{-i\oDSP}{2\MQ}
    \Big)^{-1} e^{-2iv\hspace{-0.15cm}\slash m_Q v\cdot x}(i\CDPN)
    Q^{\mp}_v,
\end{eqnarray}
where
\begin{eqnarray}
& & i\CDPN=i \VS v\cdot D +\frac{1}{2\MQ}i\DSC \Big(1-\frac{i \VS
v\cdot D }{2\MQ} \Big)^{-1} i\DSC ,
   \nonumber \\
 & & i\CDPNL = -i\oDSP +
 \frac{1}{2\MQ}   (-i\DSCXL)\Big(1-\frac{-i\oDSP}{2\MQ}\Big)^{-1}
 (-i\DSCXL)  ,\nonumber\\
&& i\DSC=i\DS - i  \VS v\cdot D , \hspace{2cm} -i \DSCXL =-i\DSXL
+i\oDSP ,
\end{eqnarray}
which is treated as HQEFT in the case that the longitudinal and
transverse residual momenta, i.e. the operators $i v\cdot D$ and
$D_{\bot}$ are at the same order of power counting in $1/m_Q$
expansions.

The heavy-light quark current $\bar{q}\Gamma Q$ with $\Gamma$
being arbitrary Dirac matrices can be expanded in powers of
$1/m_Q$ as
\begin{eqnarray}
\label{HQEFTcE} \bar{q}\Gamma Q \to e^{-im_Qv\cdot x} \bar{q}
\Big\{ \Gamma +\frac{1}{2m_Q}
  \Gamma \frac{1}{i \VS v\cdot D  }(i\DSC)^2
  +{\cal O}(\frac{1}{m^2_Q}) \Big\}\QV.
\end{eqnarray}
Here the contributions from both heavy quark and antiquark fields
have been considered.

 According to above expansions for effective
Lagrangian and effective current, one can write the matrix element
in Eq. (\ref{fdef}) as the following form in powers of $1/m_Q$,
\begin{eqnarray}
\label{matrixexp} \langle \rho|\bar{u} \gamma^{\mu} (1-\gamma^5)
b|B\rangle &=&\sqrt{\frac{m_B}{\bar{\Lambda}_B}} \Big\{ \langle
\rho| \bar{u}\gamma^{\mu} (1-\gamma^5) \QV |M_v \rangle
+\frac{1}{2m_Q}\langle \rho|\bar{u}\gamma^{\mu} (1-\gamma^5)
\frac{1}{iv\cdot D}P_+ \Big(D^2_{\bot} \nonumber\\
&+&\frac{i}{2}\sigma_{\alpha\beta} F^{\alpha\beta}\Big) \QV|M_v
\rangle +{\cal O}(1/m^2_Q) \Big\} ,
\end{eqnarray}
where $\bar{\Lambda}_B=m_B-m_b$, and $F^{\alpha\beta}$ is the
gluon field strength tensor. The effective heavy meson state $|M_v
\rangle$ satisfies the heavy quark spin-flavor symmetry. Its
normalization is
\begin{eqnarray}
\langle M_v|\QVB \gamma^\mu \QV |M_v \rangle =2\bar{\Lambda} v^\mu
\end{eqnarray}
with the binding energy $\bar{\Lambda}\equiv
  \lim_{m_Q\to \infty}  \bar{\Lambda}_M$ being heavy flavor
independent.

It should be noted that the $1/m_Q$  corrections in
Eq.(\ref{matrixexp}) include both contributions from the current
expansion (\ref{HQEFTcE}) and from the insertion of the effective
Lagrangian (\ref{HQEFTLagrangian}). In Eqs.(\ref{HQEFTcE}) and
(\ref{matrixexp}) the operator $1/(iv\cdot D)$ arises from the
contraction of effective heavy quark and antiquark fields
\cite{wwy,excit}. In the $v\cdot A=0$ gauge to be used in our
calculation, this operator is tantamount to the heavy quark
propagator.

As can be seen in Eq.(\ref{matrixexp}) that the $1/m_Q$ order
corrections to $B\to \rho l\nu$ transition are only attributed to
one kinematic operator and one chromomagnetic operator. The heavy
quark symmetry enables us to parameterize the matrix elements in
HQEFT as
\begin{eqnarray}
&&\langle \rho(p,\epsilon^*) |\bar{u}\Gamma \QV|M_v\rangle
=-Tr[\Omega(v,p)\Gamma
{\cal M}_v], \\
&&\langle \rho(p,\epsilon^*) |\bar{u}\Gamma \frac{P_+}{iv\cdot
D}D^2_{\bot}\QV
  |M_v\rangle =-Tr[\Omega_1(v,p)\Gamma {\cal M}_v], \\
&&\langle \rho(p,\epsilon^*) |\bar{u}\Gamma \frac{P_+}{iv\cdot D}
   \frac{i}{2}\sigma_{\alpha\beta}F^{\alpha\beta}\QV
  |M_v\rangle =-Tr[\Omega^{\alpha\beta}_1(v,p)\Gamma P_+
   \frac{i}{2} \sigma_{\alpha\beta} {\cal M}_v],
\end{eqnarray}
where the pseudoscalar heavy meson spin wave function $ {\cal
M}_v=-\sqrt{\bar\Lambda} P_+ \gamma^5 $ is independent of the
heavy quark flavor. $\Omega(v,p)$, $\Omega_1(v,p)$ and
$\Omega^{\alpha\beta}_1(v,p)$ can be decomposed into Lorentz
scalar functions,
\begin{eqnarray}
\label{parscalar1}
 \Omega(v,p)&=&L_1(v\cdot p) {\epsilon\hspace{-0.2cm}\slash}^*
   +L_2( v\cdot p) (v\cdot \epsilon^*) +[L_3(v\cdot p)
   {\epsilon\hspace{-0.2cm}\slash}^* +L_4(v\cdot p) (v\cdot \epsilon^* )]
   {\hat{p}\hspace{-0.2cm}\slash},   \nonumber\\
\label{parscalar2}
 \Omega_1(v,p)&=& \delta L_1(v\cdot p) {\epsilon\hspace{-0.2cm}\slash}^*
   +\delta L_2( v\cdot p) (v\cdot \epsilon^*) +[ \delta L_3(v\cdot p)
   {\epsilon\hspace{-0.2cm}\slash}^* +\delta L_4(v\cdot p) (v\cdot \epsilon^* )]
   {\hat{p}\hspace{-0.2cm}\slash},   \nonumber\\
\label{parscalar3}
 \Omega^{\alpha\beta}_1(v,p)&=& (\hat{p}^\alpha \gamma^\beta -\hat{p}^\beta \gamma^\alpha)
  \Big [{\epsilon\hspace{-0.2cm}\slash}^* (R_1+R_2 {\hat{p}\hspace{-0.2cm}\slash}
  )+(v \cdot \epsilon^* )( R_3+ R_4  {\hat{p}\hspace{-0.2cm}\slash} )  \Big
  ]\nonumber\\
  &+& (\epsilon^{*\alpha} \gamma^\beta -\epsilon^{*\beta} \gamma^\alpha
  )\Big [ R_5+ R_6 {\hat{p}\hspace{-0.2cm}\slash} \Big ] +(\epsilon^{*\alpha} \hat{p}^\beta
  -\epsilon^{*\beta} \hat{p}^\alpha ) \Big[
  R_7+R_8{\hat{p}\hspace{-0.2cm}\slash} \Big ]  \nonumber\\
  &+& i\sigma^{\alpha\beta}\Big [{\epsilon\hspace{-0.2cm}\slash}^* (R_9+ R_{10}
   {\hat{p}\hspace{-0.2cm}\slash} )+(v\cdot \epsilon^*)(R_{11}+ R_{12}
   {\hat{p}\hspace{-0.2cm}\slash} ) \Big ]
\end{eqnarray}
with $\hat{p}^\mu =p^\mu /( v\cdot p)$ and $ \xi  \equiv v\cdot p
=(m^2_B+m^2_\rho-q^2)/{2m_B} $.

Eqs.(\ref{matrixexp})-(\ref{parscalar3}) yield
\begin{eqnarray}
\label{matrixinfac} \langle \rho(p,\epsilon^*)|\bar{u}\gamma^\mu
(1-\gamma^5) b|B(p_B)\rangle
 &=& -2i \sqrt{\frac{m_B \bar\Lambda}{\bar{\Lambda}_B}} \Big\{
 i L'_3 \epsilon^{\mu\nu\alpha\beta} \epsilon_\nu \hat{p}_\alpha
 v_\beta  + (L'_1+L'_3)\epsilon^{*\mu}  \nonumber \\
 &-& (L'_3-L'_4) (v\cdot \epsilon^*) \hat{p}^\mu
 -L'_2 (v\cdot \epsilon^*) v^\mu +{\cal O}(1/m^2_Q)
 \Big\}
\end{eqnarray}
with
\begin{eqnarray}
L'_i&=& L_i+\frac{1}{2m_Q} \delta L'_i
   = L_i+ \frac{1}{2m_Q}( \delta L_i +R'_i) , \nonumber\\
R'_1&=& -2R_1-2R_5 +R_7 -3R_9 +(2R_2 -R_8) \hat{p}^2 ,  \nonumber\\
R'_2&=& -2R_3 -2R_5 -3R_{11}-(2R_4 +R_8) \hat{p}^2  , \nonumber\\
R'_3&=& 2R_1 -2R_2 -2R_6 -R_7 +R_8 -3R_{10} , \nonumber\\
R'_4&=& -2R_6 -R_7 -2R_3 -2R_4 -3R_{12}  . \nonumber
\end{eqnarray}
Comparison between Eqs.(\ref{fdef}) and (\ref{matrixinfac}) gives
relations between the form factors $A_i (i=1,2,3)$, $V$ and the
universal wave functions,
\begin{eqnarray}
\label{AandL} A_1(q^2)&=&\frac{2}{m_B+m_\rho} \sqrt{\frac{m_B
\bar\Lambda}{\bar\Lambda_B}}
    \{ L'_1(v\cdot p)+L'_3(v\cdot p) \}  +\cdots ;  \nonumber\\
A_2(q^2)&=&2 (m_B+m_\rho) \sqrt{\frac{m_B
\bar\Lambda}{\bar\Lambda_B}}
    \{\frac{L'_2(v\cdot p)}{2 m^2_B} +\frac{L'_3(v\cdot p)
    -L'_4(v\cdot p)}{2m_B (v\cdot p)} \}  +\cdots ;  \nonumber\\
A_3(q^2)&=&2 (m_B+m_\rho) \sqrt{\frac{m_B
\bar\Lambda}{\bar\Lambda_B}}
    \{ \frac{L'_2(v\cdot p)}{2 m^2_B} -\frac{L'_3(v\cdot p)
    -L'_4(v\cdot p)}{2m_B (v\cdot p)} \}   + \cdots ; \nonumber\\
V(q^2)&=&\sqrt{\frac{m_B \bar\Lambda}{\bar\Lambda_B}}
    \frac{m_B+m_\rho}{m_B (v\cdot p) } L'_3(v\cdot p) +\cdots ,
\end{eqnarray}
where the dots denote higher order $1/m_Q$ contributions not to be
taken into account in the following calculations.

\section{Light cone sum rules in HQEFT}\label{sumrule}

For the derivation of the $1/m_Q$ order corrections to $B\to \rho
l\nu $ decay, we consider the two-point correlation function
\begin{eqnarray}
\label{correlator} F^\mu&=&i\int d^4x e^{-i(p_B-m_Qv)\cdot
x}\langle \rho(p,\epsilon^*)|T\Big\{\bar{u}(0)\gamma^\mu
(1-\gamma^5)
 \frac{1}{iv\cdot D}P_+ \Big(D^2_\bot +\frac{i}{2} \sigma_{\alpha\beta}
 F^{\alpha\beta}\Big)\QV(0), \nonumber\\
&& \QVB(x)i\gamma^5 d(x) \Big\}|0 \rangle
\end{eqnarray}
where $\QVB(x) i\gamma^5 d(x)$ is the interpolating current for
$B$ meson. Inserting between the two currents in
Eq.(\ref{correlator}) a complete set of intermediate states with
the $B$ meson quantum number, one gets
\begin{eqnarray}
\label{correlatorhad} && \frac{m_B \bar{\Lambda}}{m_Q
\bar{\Lambda}_B} \frac{2iF}{2\bar{\Lambda}_B- \omega} \Big\{
\delta L'_3 \epsilon^{\mu\nu\alpha\beta} \epsilon^*_\nu
\hat{p}_\alpha v_\beta +i(\delta L'_3-\delta L'_4) (v\cdot
\epsilon^*) \hat{p}^\mu -i(\delta L'_1+\delta L'_3)\epsilon^{*\mu}
+i \delta L'_2 (v\cdot \epsilon^*) v^\mu \Big\} \nonumber\\
&& +\int^\infty_{s_0} ds \frac{ \rho_(\xi, s)
 } {s- \omega}
 +subtraction
\end{eqnarray}
with $\xi \equiv v\cdot p$ and $\omega \equiv 2v\cdot k$, where
$k=p_B-m_Q v$ is the residual momentum of the bottom quark. The
second term in (\ref{correlatorhad}) represents the higher
resonance contributions. $F$ is the decay constant of $B$ meson at
the leading order of $1/m_Q$ expansion, defined by \cite{ww}
\begin{equation}
\langle 0|\bar{q}\Gamma \QV|B_v \rangle =\frac{F}{2}Tr[\Gamma {\cal M}_v].
\end{equation}
In deep Euclidean region the correlator (\ref{correlator}) can be
calculated in effective field theory. The result can be written
also as an integral over a theoretic spectral density,
\begin{equation}
\label{correlatorth} \int^{\infty}_{0} ds \frac{\rho_{th}( \xi,s)}
{s-\omega}+subtraction .
\end{equation}

The standard treatment of sum rule is to assume the quark-hadron
duality, and to equal the hadronic representation
(\ref{correlatorhad}) and the theoretic one (\ref{correlatorth}),
which provides an equation:
\begin{eqnarray}
\label{correlatoreq} && \frac{m_B \bar{\Lambda}}{m_Q
\bar{\Lambda}_B} \frac{2iF}{2\bar{\Lambda}_B- \omega} \Big\{
\delta L'_3 \epsilon^{\mu\nu\alpha\beta} \epsilon^*_\nu
\hat{p}_\alpha v_\beta +i(\delta L'_3-\delta L'_4) (v\cdot
\epsilon^*) \hat{p}^\mu -i(\delta L'_1+\delta L'_3)\epsilon^{*\mu}
+i \delta L'_2 (v\cdot \epsilon^*) v^\mu \Big\} \nonumber\\
&& +\int^\infty_{s_0} ds \frac{ \rho_(\xi,s)
 } {s- \omega}
= \int^{\infty}_{0} ds \frac{\rho_{th}( \xi,s)}
{s-\omega}+subtraction .
\end{eqnarray}
To ensure the reliability of sum rule estimates, one should
enhance the importance of the ground state contribution, suppress
higher order nonperturbative contributions and remove the
$subtraction$. These can be achieved by performing the Borel
transformation
\[
\hat{B}^{(\omega)}_{T}\equiv \lim_{
{\tiny \begin{array}{c}
-\omega,n\to \infty \\
  -\omega/n=T  \end{array} }  }
 \frac{(-\omega)^{n+1}}{n!}(\frac{d}{d\omega})^n
\]
to both sides of the equation (\ref{correlatoreq}). With using the
formulae
\begin{eqnarray}
\label{borelfeature}
\hat{B}^{(\omega)}_T \frac{1}{s-\omega}=e^{-s/T}, \hspace{2cm}
\hat{B}^{(\omega)}_{T} e^{\lambda \omega}=\delta(\lambda-\frac{1}{T}),
\end{eqnarray}
one gets
\begin{eqnarray}
\label{corequ} && 2iF \{ \delta L'_3 \epsilon^{\mu\nu\alpha\beta}
\epsilon^*_\nu \hat{p}_\alpha v_\beta -i(\delta L'_1+\delta
L'_3)\epsilon^{*\mu} +i(\delta L'_3-\delta L'_4) (v\cdot
\epsilon^*) \hat{p}^\mu +i \delta L'_2 (v\cdot \epsilon^*)
v^\mu \} e^{-2 \bar{\Lambda}_B/T} \nonumber\\
&&= \int^{s_0}_{0} ds e^{-s/T} \rho_(\xi,s) ,
\end{eqnarray}
where the spectral density $\rho(\xi, s)$ can be derived via
double Borel transformations,
\begin{equation}
\label{Borel2kernal} \rho(\xi,s) =\hat{B}^{(-1/T)}_{1/s}
\hat{B}^{(\omega)}_{T}F^\mu(\xi,\omega)  .
\end{equation}

In calculating the three-point function (\ref{correlator}), one
may represent the nonperturbative contributions embeded in the
hadronic matrix element in terms of light cone wave functions.
Among them are the two-particle distribution functions and the
three-particle ones. However, if we restrict our calculation to
the lowest twist (twist 2) level, the three-particle functions do
not contribute. As a result, the chromomagnetic operator in
Eq.(\ref{correlator}) could be neglected in the lowest twist
approximation. The leading twist distribution functions are
defined by \cite{ar,pvda,pvmisu,pvesrb}
\begin{eqnarray}
\label{wfdef} <\rho(p,\epsilon^*)|\bar{u}(0) \sigma_{\mu\nu}
d(x)|0>&=&-i f^\bot_\rho
  (\epsilon^*_\mu p_\nu - \epsilon^*_\nu p_\mu)
\int^1_0 du e^{iup\cdot x} \phi_\bot (u)  ,\nonumber\\
<\rho(p,\epsilon^*)|\bar{u}(0) \gamma_\mu d(x)|0>&=& f_\rho m_\rho
p_\mu
  \frac{\epsilon^*\cdot x}{p\cdot x} \int^1_0 du e^{iup\cdot x} \phi_{\|}(u) \nonumber\\
  &+& f_\rho m_\rho (\epsilon^*_\mu-p_\mu \frac{\epsilon^* \cdot x}{p\cdot x})
  \int^1_0 du e^{iup\cdot x} g^{(v)}_\bot (u), \nonumber\\
<\rho(p,\epsilon^*)|\bar{u}(0) \gamma_\mu \gamma_5 d(x)|0>&=&
  \frac{1}{4} f_\rho m_\rho \epsilon_{\mu\nu\alpha\beta} \epsilon^{*\nu} p^\alpha x^\beta
  \int^1_0 du e^{iup\cdot x} g^{(a)}_\bot (u)
\end{eqnarray}
with $\phi_{\bot,\|}$, and $g^{(v,a)}_{\bot}$ being functions with
nonperturbative nature.

Then the effective heavy quark fields $\QV(x_1)\QVB(x_2)$ can be
contracted into a propagator of heavy quark, $P_+
\int^{\infty}_{0}dt \delta(x_1-x_2-vt)$. In the lowest twist
approximation, only the kinematic operator contributes to the
$1/m_Q$ order corrections to $B\to \rho l\nu$ decay. At $v\cdot
A=0$ gauge we used, the correlation function (\ref{correlator})
simplifies as
\begin{eqnarray}
\label{midcal} && i \int d^4x \int d^4y \int^{\infty}_0 d l
\int^{\infty}_0 dt e^{-ik\cdot x} \langle
\rho(p,\epsilon^*)|\bar{u}(0)\gamma^\mu (1-\gamma^5)
P_+ \delta(-y-vl) \nonumber\\
&&\hspace{2cm} \times  \Big[\partial^2_{(y)} -v_\alpha v_\beta
\partial^\alpha_{(y)} \partial^\beta_{(y)} -\partial^\alpha_{(y)}
A_\alpha (y) -A_\alpha (y) \partial^\alpha_{(y)}
%+v_\alpha v_\beta \partial^\alpha_{(y)} A^\beta (y)
\nonumber\\
&&\hspace{2cm}
%+v_\alpha v_\beta A^\alpha (y) \partial^\beta_{(y)}
 +\frac{i}{2}
\sigma_{\alpha \beta} F^{\alpha \beta} (y) +A_\alpha (y) A^\alpha
(y)
%-(v \cdot A)^2
\Big]
  P_+ \delta(y-x-vt) \gamma^5 d(x)|0 \rangle  \nonumber\\
&& \rightarrow  i \int d^4x \int d^4y \int^{\infty}_0 d l
\int^{\infty}_0 dt e^{-ik\cdot x} \delta (-y-v l) \Big[(
\frac{\partial^2}{\partial y^2}-v^\alpha v^\beta
\frac{\partial^2}{\partial y^\alpha \partial y^\beta}) \delta
(y-x-vt) \Big ] \nonumber\\
&& \hspace{2cm} \langle \rho(p,\epsilon^*)|\bar{u}(0) \frac{1}{2}
\Big [ \gamma^\mu \gamma^5 -\gamma^\mu - \sigma^{\alpha\beta}
(iv_\beta g_{\alpha\mu} +\frac{1}{2} \epsilon_{\mu\nu\alpha\beta}
v_\nu ) \Big ]
 d(x)|0 \rangle
\end{eqnarray}
with $\partial^\alpha_{(y)} \equiv \partial / (\partial
y_\alpha)$. The final formula in (\ref{midcal}) includes only the
terms related to the two-particle distribution functions
(\ref{wfdef}).

Now the transition matrix element can be evaluated through the
distribution functions defined in (\ref{wfdef}). Using Eqs.
(\ref{borelfeature}) and (\ref{wfdef}), the spectral function is
found to be
\begin{eqnarray}
\label{srrhoa}
 \rho(\xi,s)&=& \hat{B}^{(-1/T)}_{1/s} \hat{B}^{(\omega)}_{T}F^\mu
 =\frac{1}{2\xi} \Big\{ i\epsilon^{\mu\nu\alpha\beta}
 \epsilon^*_\nu p_\alpha v_\beta \Big [ -\frac{1}{4 \xi^2}f_\rho
 m^3_\rho (u^2 g^{(a)}_\bot)''+f^{\bot}_\rho m^2_\rho
 \frac{1}{\xi} (u^2 \phi_\bot)'  \nonumber\\
 &+& \frac{1}{4}f_\rho m_\rho
 (u^2 g^{(a)}_\bot)''- \frac{3}{4}f_\rho m^2_\rho (u
 g^{(a)}_\bot)' \Big ]+\epsilon^{*\mu} \Big [ f_\rho m^3_\rho
 \frac{1}{\xi}(u^2 g^{(v)}_\bot)'-2f^\bot_\rho m^2_\rho u
 \phi_\bot  \nonumber\\
 &-& f_\rho m_\rho \xi (u^2 g^{(v)}_\bot)'-2 f^\bot_\rho
 \xi^2 u \phi_\bot+4f^\bot_\rho \xi^2 u \phi_\bot \Big ]
 +(v\cdot \epsilon^*) p^\mu \Big [-\frac{1}{\xi^2}f_\rho m^3_\rho
 (u^2 g^{(v)}_\bot)'  \nonumber\\
 &+& \frac{1}{\xi^2} f_\rho m^3_\rho (u^2
 \phi_{\|})' -\frac{2}{\xi^2} f_\rho m^3_\rho u g^{(v)}_\bot
 +\frac{2}{\xi^2} f_\rho m^3_\rho u \phi_{\|} -\frac{2}{\xi^3}
 f_\rho m^3_\rho \xi G^{(v)}_{\bot} +\frac{2}{\xi^3} f_\rho
 m^3_\rho \xi \Phi_{\|}  \nonumber\\
 &+& f_\rho m_\rho (u^2 g^{(v)}_\bot)' -f_\rho
 m_\rho (u^2 \phi_{\|})' \Big ]
+(v\cdot \epsilon^*) v^\mu \Big [ 2 f^\bot_\rho m^2_\rho u
\phi_\bot -2 f^\bot_\rho m^2_\rho (u^2 \phi_\bot)'
\nonumber\\
&-& 4f^\bot_\rho \xi^2 u \phi_\bot +2 f^\bot_\rho \xi^2 (u^2
\phi_\bot)' \Big ] \Big \}_{u \to \frac{s}{2\xi}}    ,
\end{eqnarray}
where ¡°$'$¡± denotes derivative with respect to the variable $u$,
while $G^{(v)}_\bot(u)$ and $\Phi_{\|}(u)$ are functions related
to $g^{(v)}_\bot$ and $\phi_{\|}$ by $\frac{\partial}{\partial u }
G^{(v)}_\bot (u) =g^{(v)}_\bot (u)$, $\frac{\partial}{\partial u}
\Phi_{\|}(u)=\phi_{\|}(u) $. The detailed procedure in deriving
Eq.(\ref{srrhoa})are similar to those in
Refs.\cite{bpi,brho,bpinlo}.

\section{Numerical analysis}\label{discussion}

$\phi_{\bot}$ and $\phi_{\|}$ are the lowest twist distributions
in the fraction of total momentum carried by the quark in
transversely and longitudinally polarized mesons. They can be
expanded in Gegenbauer polynomials $C^{3/2}_n(x)$ whose
coefficients are renormalized multiplicatively. With the scale
dependence explicitly, one has \cite{pvda}
\begin{eqnarray}
\label{rhowfmu} \phi_{\bot(\|)}(u,\mu)&=&6u(1-u)[
1+\sum_{n=2,4,\cdots} a^{\bot(\|)}_n(\mu)
   C^{3/2}_n(2u-1)  ], \nonumber\\
a^{\bot(\|)}_n(\mu)&=&a^{\bot(\|)}_n(\mu_0) (\frac{\alpha_s(\mu)}
   {\alpha_s(\mu_0)})^{(\gamma^{\bot (\|)}_n-\gamma^{\bot (\|)}_0)/(2\beta_0)},
\end{eqnarray}
where $\beta_0=11-(2/3)n_f$, and $\gamma^{\|}_n$,
$\gamma^{\bot}_n$ are the one loop anomalous dimensions
\cite{dfmn,mm}. The nonperturbative parameters $a^\bot_n$ and
$a^{\|}_n$ have been obtained in \cite{pvda} with the values
\begin{eqnarray}
\label{para} a^\bot_2(1\mbox{GeV})=0.2\pm 0.1, \;\;\;
a^{\|}_2(1\mbox{GeV})=0.18\pm 0.10
\end{eqnarray}
and $a^{\bot(\|)}_n=0$ for $n\neq 2$.

The functions $g^{(v)}_\bot$ and $g^{(a)}_\bot$ describe
transverse polarizations of quarks in the longitudinally polarized
mesons. As in Ref.\cite{pvesrb} they are parameterized as
\begin{eqnarray}
\label{gcva} g^{(v)}_\bot (u,\mu)&=& \frac{3}{4} \Big (1+(2u-1)^2
\Big )+\frac{3}{2} a^{\|}_1(\mu) (2u-1)^3+(\frac{3}{7}
a^{\|}_2(\mu)+5
\xi_3(\mu))\Big (3(2u-1)^2-1 \Big ) \nonumber \\
&+& \Big [ \frac{9}{112}a^{\|}_2(\mu) +\frac{15}{64} \xi_3(\mu) (3
\omega^V_3(\mu) -\omega^A_3(\mu)) \Big ] \Big ( 3-30
(2u-1)^2+35 (2u-1)^4 \Big ), \nonumber\\
g^{(a)}_\bot (u,\mu)&=& 6u(1-u) \Big [ 1+a^{\|}_1(\mu) (2u-1)+\Big
( \frac{1}{4} a^{\|}_2(\mu) +\frac{5}{3} \xi_3(\mu)
(1-\frac{3}{16} \omega^A_3(\mu) +\frac{9}{16} \omega^V_3(\mu) )
\Big ) \nonumber\\
&\times & \Big ( 5(2u-1)^2-1 \Big )\Big ]
\end{eqnarray}

All the nonperturbative parameters in Eqs.(\ref{gcva}) have been
estimated in Ref.\cite{pvesrb}. The asymptotic form of these
factors and the renormalization scale dependence are given by
perturbative QCD \cite{bf,va}. As in Refs.\cite{brho}, the typical
virtuality of the bottom quark
\begin{equation}
\label{mub} \mu_b \sim \sqrt{m^2_B-m^2_b} \approx 2.4\mbox{GeV},
\end{equation}
is used for the energy scale for the current calculation.

The values of the hadron quantities $f_\rho$, $f^\bot_\rho$,
$\bar\Lambda_B$, $\bar\Lambda$ and $F$ have been extracted in the
previous work (see, e.g., \cite{pvda,ww,mab,lm}). For consistency,
here we use for them the same values as in Ref.\cite{brho},i.e.,
\begin{eqnarray}
\label{paravalue} && f_{\rho^{\pm}}=(195\pm 7)\mbox{MeV}, \;\;\;
f_{\rho^0}=(216\pm 5)\mbox{MeV}, \;\;\;
f^\bot_\rho=(160 \pm 10 )\mbox{MeV}, \nonumber \\
&& \bar\Lambda_B \approx \bar\Lambda=0.53 \mbox{GeV}, \;\;\;\;
F=(0.30 \pm 0.06)\mbox{GeV}^{3/2}.
\end{eqnarray}

$\delta L_i$ as functions of $\xi$, $T$ and $s_0$ can be derived
from Eqs.(\ref{corequ}) and (\ref{srrhoa}). Fig.1 shows the
variation of $\delta L_i$ as functions of the Borel parameter $T$
at $v\cdot p=2.5$ GeV. The curves in each figure correspond to
different values adopted for the threshold $s_0$.

The rule of LCSR method is to determine $s_0$ from the stability
of relevant curves in the reliable region of $T$, where both the
higher nonperturbative corrections and the contributions from
excited and continuum states should not be large. In the current
case, we focus on the region around $T=1.5-2$GeV. As shown in
Fig.1, $\delta L_i$ are found to be stable with respect to the
Borel parameter $T$. In Ref.\cite{brho} the threshold $s_0=2.1 \pm
0.6$GeV is adopted in evaluating the leading order wave functions
$L_i$. In calculating the decay width we will use for $\delta L_i$
the same threshold values as those for the leading order wave
function $L_i$, i.e., $s_0 \approx 2.1$GeV.

With (\ref{AandL}), the form factors $A_1$, $A_2$, $A_3$ and $V$
with including $1/m_Q$ order corrections can be calculated. It is
convenient to represent each of these form factors in terms of
three parameters as
\begin{eqnarray}
\label{fitform} F(q^2)=\frac{F(0)}{1-a_F q^2/m^2_B+b_F
(q^2/m^2_B)^2},
\end{eqnarray}
where $F(q^2)$ can be any one of $A_1(q^2)$, $A_2(q^2)$,
$A_3(q^2)$ and $V(q^2)$. The parameters $F(0)$, $a_F$ and $b_F$
presented in table 1 are fitted from the the LCSR results at
$s_0=2.1$GeV. Fig.2 shows the form factors as functions of the
momentum transfer, where the dashed curves are for the leading
order results while the solid ones for the results with the
$1/m_Q$ order corrections included.
\begin{center}
\begin{tabular}{c|c|c|c|c}
\hline \hline
 & \hspace{0.9cm} $F(0)$ \hspace{0.9cm} & \hspace{0.9cm} $a_F$ \hspace{0.9cm} &
\hspace{0.9cm} $b_F$ \hspace{0.9cm} & \hspace{0.5cm}  \hspace{0.9cm} \\
\hline $A_1$ & $0.26$   & $0.37 $
     & $-0.19$ & LO\\
\cline{2-5}
  & $0.27 $  & $0.32 $ & $-0.19 $
     & NLO \\
\hline $A_2$ & $0.26 $   & $1.11  $
     & $0.26 $ & LO\\
\cline{2-5}
  & $0.26 $  & $1.15 $ & $0.30 $
     & NLO \\
\hline $A_3$ & $-0.26$   & $1.12 $
     & $0.26$ & LO\\
\cline{2-5}
  & $-0.26 $  & $1.11 $ & $0.25 $
     & NLO \\
\hline $V$ & $0.32 $   & $1.24  $
     & $0.29 $ & LO\\
\cline{2-5}
  & $0.31 $  & $1.26 $ & $0.32 $
     & NLO \\
\hline \hline
\end{tabular}
\end{center}

\vspace{0cm} \centerline{
\parbox{13cm}{
\small \baselineskip=1.0pt Table 1. Results of LCSR calculations
up to leading (LO) and next leading order (NLO) in HQEFT. The
leading order results are obtained in Ref.\cite{bpi}. } }

\vspace{0.5cm}

The differential decay width of $B\to \rho l\nu $ with the lepton
mass neglected is
\begin{eqnarray}
\label{gammaq2} \frac{d\Gamma}{dq^2}=\frac{G^2_F |V_{ub}|^2}{192
\pi^3 m^3_B}
  \lambda^{1/2} q^2 (H^2_0+H^2_+ + H^2_-)
\end{eqnarray}
with the helicity amplitudes
\begin{eqnarray}
\label{heliampli}
H_{\pm}&=&(m_B+m_\rho) A_1(q^2) \mp \frac{\lambda^{1/2}}{m_B+m_\rho} V(q^2),\nonumber\\
H_0&=&\frac{1}{2m_\rho \sqrt{q^2}} \{ (m^2_B-m^2_\rho-q^2)
(m_B+m_\rho) A_1(q^2)
    -\frac{\lambda}{m_B+m_\rho}A_2(q^2)  \}
\end{eqnarray}
and
\begin{eqnarray}
\label{ladd} \lambda \equiv (m^2_B+m^2_\rho-q^2)^2-4m^2_B
m^2_\rho.
\end{eqnarray}
The total width of $B \to \rho l \nu$ can be obtained by
integrating (\ref{gammaq2}) over the whole accessible region of
$q^2$. We get
\begin{eqnarray}
\label{width} \Gamma(B \to \rho l \nu)=(13.6 \pm 4.0 ) |V_{ub}|^2
\mbox{ps}^{-1},
\end{eqnarray}
where the error results from the variation of the threshold energy
$s_0=1.5-2.7$GeV.

The branching fraction of $B^0 \to \rho^- l^+ \nu$ is measured to
be $\mbox{Br}(B^0 \to \rho^- l^+ \nu)=(2.6\pm 0.7) \times 10^{-4}$
\cite{pdg}. This and the world average of the $\mbox{B}^0$
lifetime \cite{pdg} $\tau_{\tiny{\mbox{B}^0}}=1.536\pm 0.014
\;\mbox{ps}$ yields
\begin{eqnarray}
\label{CLEO} \Gamma(B^0\to \rho^- l^+ \nu)=(1.69 \pm 0.47 )\times
10^{-4} \mbox{ps}^{-1}.
\end{eqnarray}
$|V_{ub}|$ is then extracted from Eqs.(\ref{width}) and
(\ref{CLEO}). It is
\begin{eqnarray}
|V_{ub}|=(3.53 \pm 0.49 \pm 0.52)\times 10^{-3},
\end{eqnarray}
where the first and second errors correspond to the experimental
and theoretical uncertainties, respectively. This value may be
compared with the ones previously obtained
\cite{bpi,brho,bpinlo,yww,wy}. From the exclusive semileptonic
decays $B\to \pi(\rho) l\nu$, we then have
\begin{eqnarray}
|V_{ub}|&=&( 3.4 \pm 0.5 \pm 0.5 )\times 10^{-3} \;\; (\mbox{$B\to
\pi
l\nu$, LO}) \nonumber \\
|V_{ub}|&=&( 3.2 \pm 0.5 \pm 0.4 )\times 10^{-3} \;\; (\mbox{$B\to
\pi
l\nu$, to NLO}) \nonumber \\
|V_{ub}|&=&( 3.7 \pm 0.6 \pm 0.7 )\times 10^{-3} \;\; (\mbox{$B\to
\rho
l\nu$, LO}) \nonumber \\
|V_{ub}|&=&( 3.5 \pm 0.5 \pm 0.5 )\times 10^{-3} \;\; (\mbox{$B\to
\rho l\nu$, to NLO}) \nonumber \\
|V_{ub}|&=&( 3.5 \pm 0.6 \pm 0.1 )\times 10^{-3} \;\; (\mbox{$B$
inclusive semileptonic decays}) \nonumber
\end{eqnarray}

%\section{summary}\label{summary}

As a summary, we have studied $B\to \rho l\nu$ decay up to the
$1/m_Q$ order corrections in HQEFT. In HQEFT, $1/m_Q$ order
corrections from the effective current and from effective
Lagrangian are given by the same operator forms, which simplifies
the structure of transition matrix elements. These $1/m_Q$ order
contributions have been calculated using light cone sum rules with
considering the lowest twist distribution functions. Numerically,
the $1/m_Q$ order wave functions give only corrections lower than
10\% to the transition form factors. Similar to the $B\to \pi l\nu
$ case, the correction indicates a slightly smaller value of the
CKM matrix element $|V_{ub}|$. The discussion concerning $1/m_Q$
order corrections to $B\to \rho l\nu$ decay in this paper is also
applicable to other heavy to light vector meson decays.

\acknowledgments

This work was supported in part by the key projects of
National Science Foundation of China (NSFC)
and Chinese Academy of Sciences,
and by the BEPC National Lab Opening Project.

\newpage
\centerline{\large{FIGURES}}

\newcommand{\PICLI}[2]
{
\begin{center}
\begin{picture}(500,120)(0,0)
\put(0,20){ \epsfxsize=7cm \epsfysize=5cm \epsffile{#1} }
\put(115,10){\makebox(0,0){#2}}
\end{picture}
\end{center}
}

\newcommand{\PICRI}[2]
{
\begin{center}
\begin{picture}(300,0)(0,0)
%\put(160,50){ \epsfxsize=7cm \epsfysize=5cm \epsffile{#1} }
\put(160,45){ \epsfxsize=7cm \epsfysize=5cm \epsffile{#1} }
\put(275,34){\makebox(0,0){#2}}
\end{picture}
\end{center}
}

%========================

\small \mbox{} {\vspace{2cm}}

\PICLI{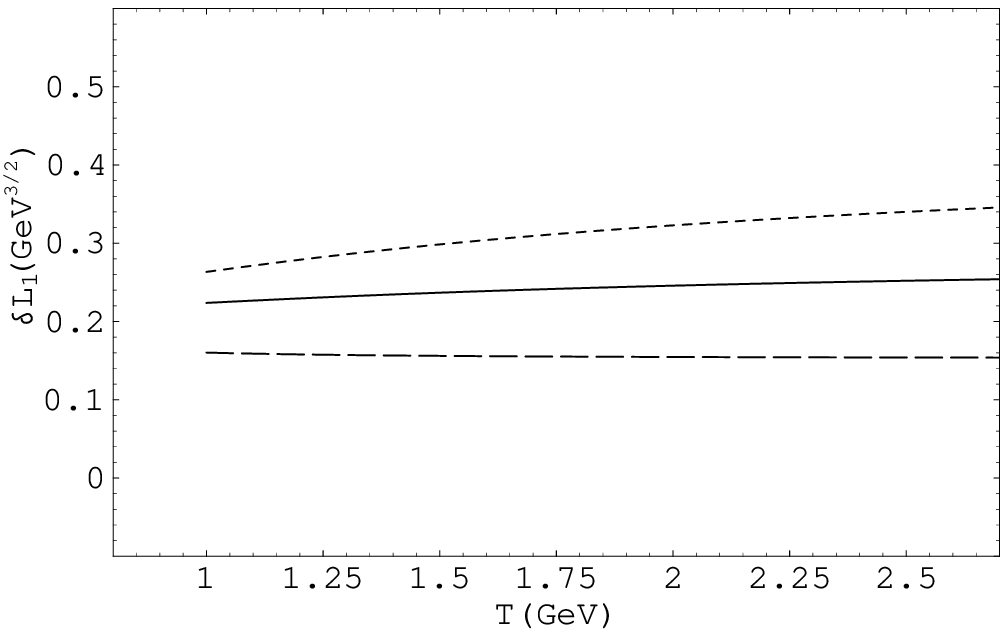}{(a)}

\PICRI{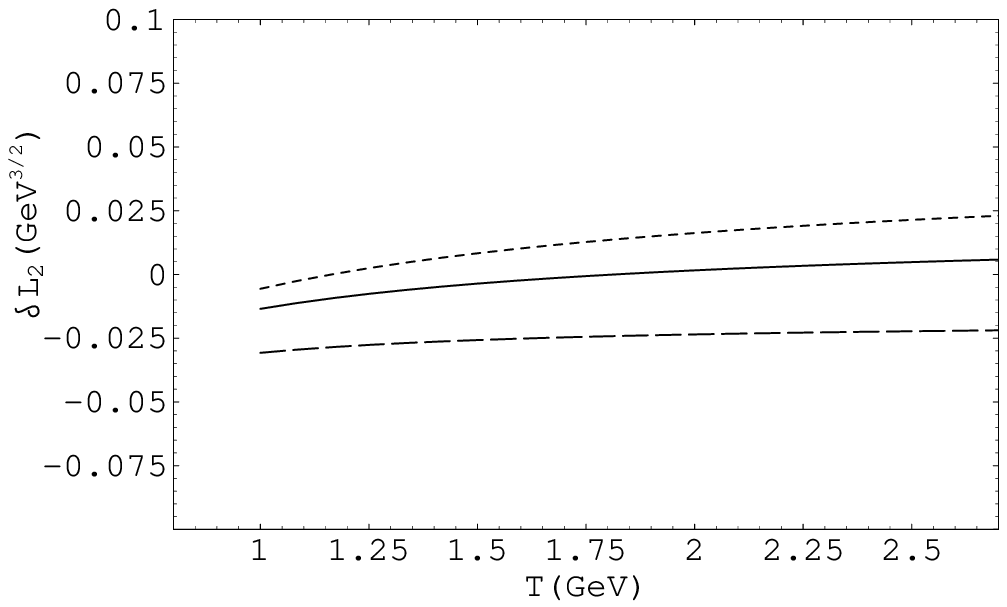}{(b)}

\vspace{1cm}

\PICLI{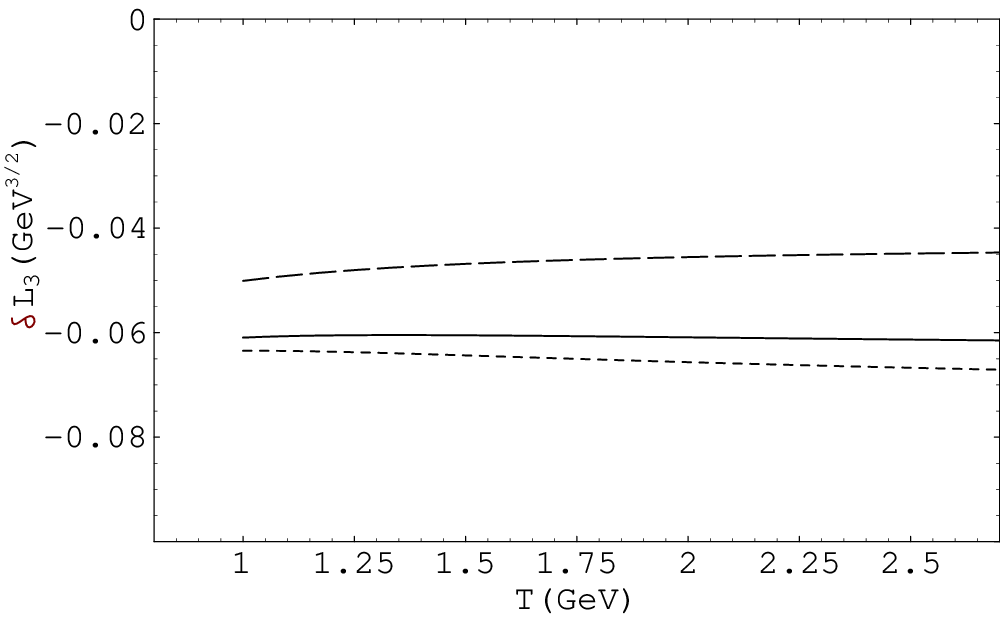}{(c)}

\PICRI{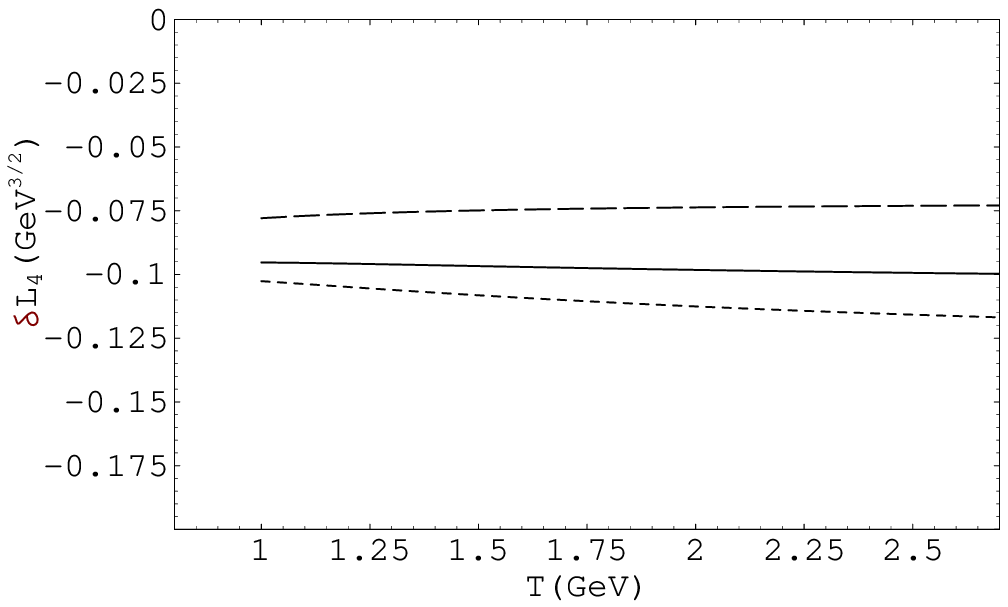}{(d)}

\vspace{-1cm}
\centerline{
\parbox{12cm}{
\small \baselineskip=1.0pt Fig.1. Variation of $1/m_Q$ order wave
functions $\delta L_i (i=1,2,3,4)$ with respect to the Borel
parameter $T$ at $\xi=v\cdot p=2.5$ GeV. The dashed, solid and
dotted curves correspond to the thresholds $s_0=$1.5, 2.1 and 2.7
GeV respectively. }}

\newpage

\small \mbox{} {\vspace{3cm}}

\PICLI{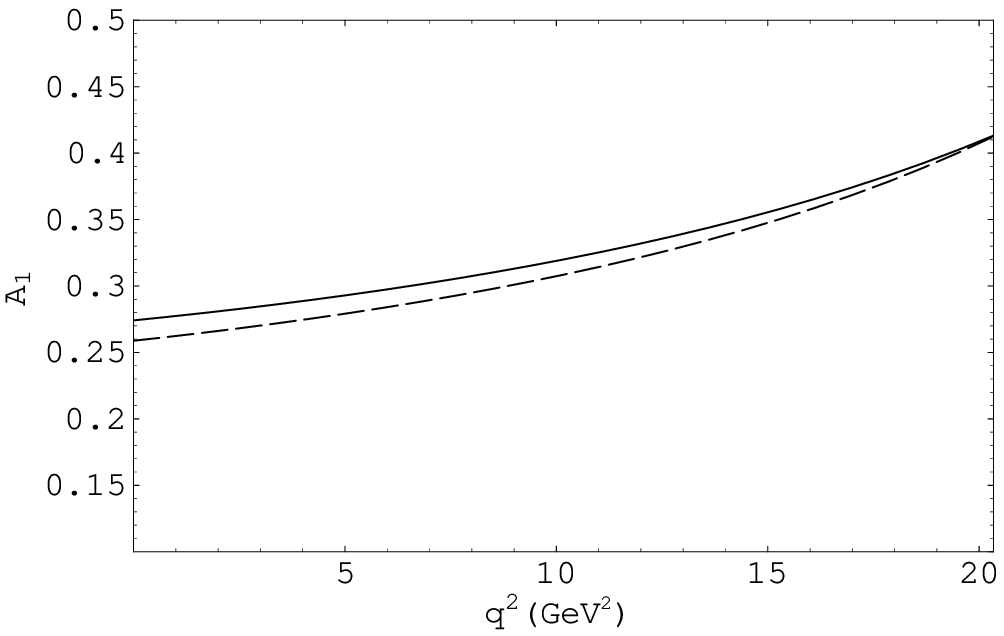}{(a)}

\PICRI{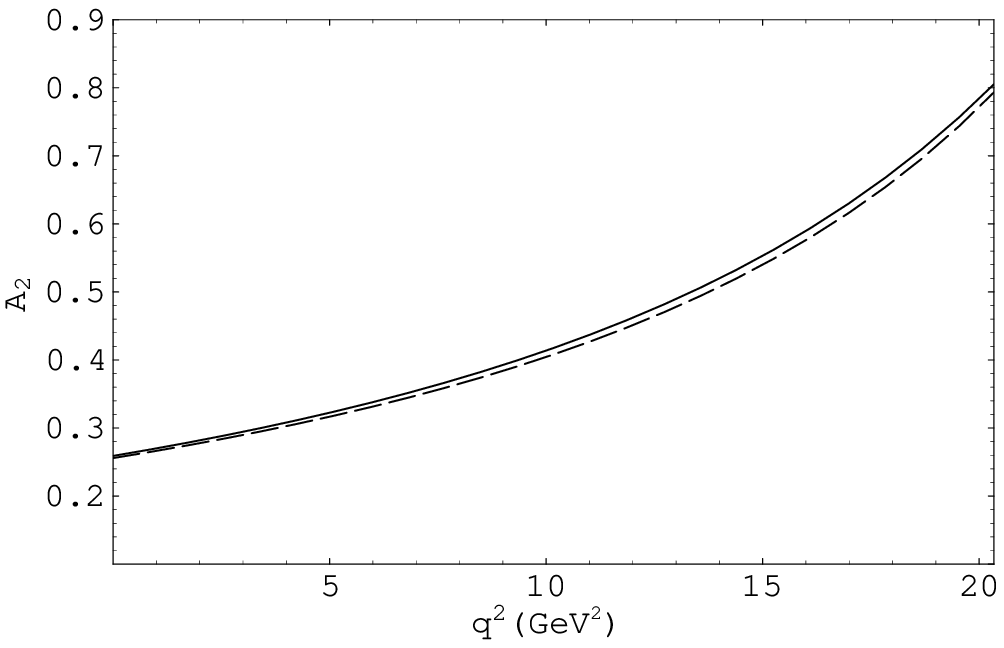}{(b)}

\vspace{1cm}

\PICLI{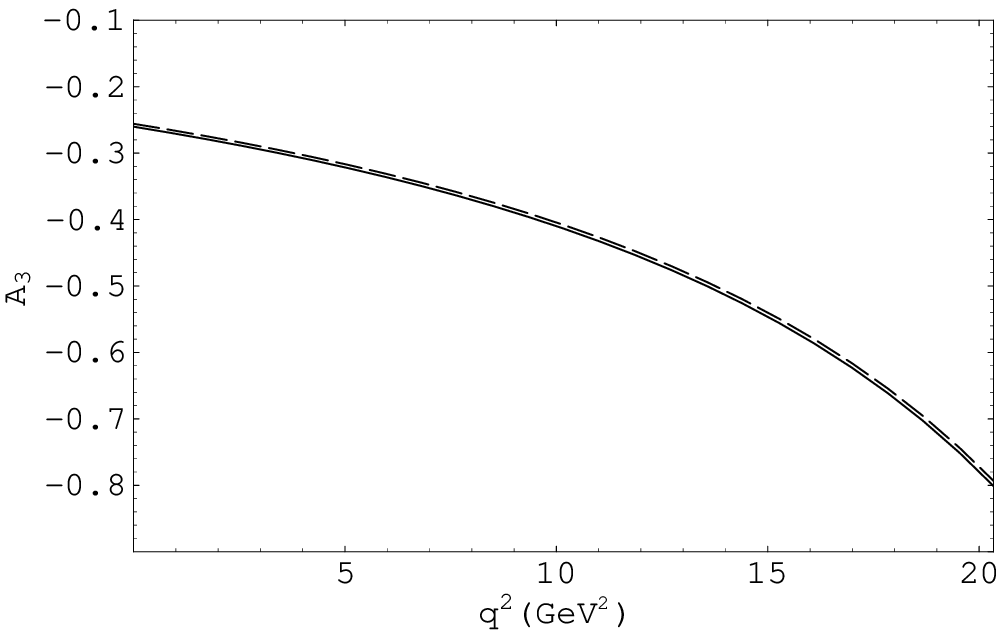}{(c)}

\PICRI{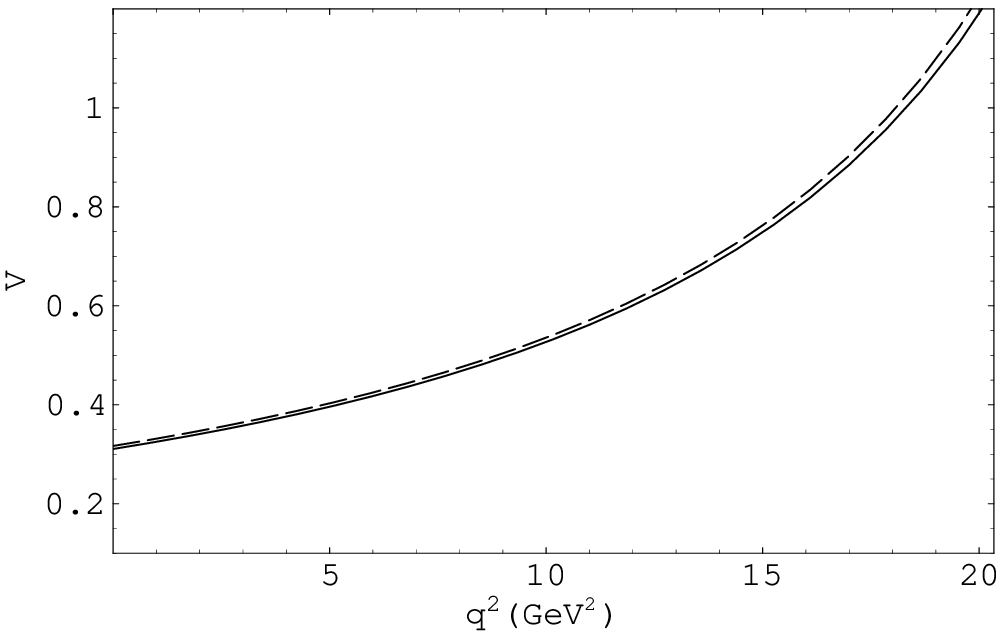}{(d)}

\vspace{-1cm} \centerline{
\parbox{12cm}{
\small \baselineskip=1.0pt Fig.2. Form factors $A_i (i=1,2,3)$ and
$V$ obtained from light cone sum rules in HQEFT. The dashed curves
are the leading order results in HQEFT \cite{bpi,h2l}, while the
solid curves are the results with including $1/m_Q$ order
correction.}}

\end{document}